\documentclass[11pt]{article}
\evensidemargin=\oddsidemargin

\usepackage{graphicx}
\usepackage{epsfig}    
\usepackage{dcolumn}
\usepackage{bm}
\usepackage{amssymb}
\usepackage{epsfig}    
\usepackage{here}

\renewcommand{\thefootnote}{\fnsymbol{footnote}}

\def\beqn{\begin{eqnarray}}
\def\eeqn{\end{eqnarray}}
\relax
\newcommand{\ba}[1]{\begin{array}{#1}}
\def\ea{\end{array}}

\def\beq{\begin{equation}}
\def\eeq{\end{equation}}
\def\bea{\begin{array}}
\def\eea{\end{array}}

\def\to{\rightarrow}

\def\dis{\displaystyle}
\def\f{\frac}
\def\[{\left[}
\def\]{\right]}
\def\({\left(}
\def\){\right)}

\def\la{{\lambda}}
\def\ep{{\epsilon}}
\def\qs{{\sqrt{2}}}

\def\Ah{{\widehat{A}}}
\def\ms{{\widetilde{m}}}
\def\sm0{{\widetilde{m}_0}}
\def\sB{{\sin\beta}}

\def\tanB{{\tan\beta}}
\def\cotB{{\cot\beta}}
\def\sM{{\widetilde{\cal M}}}
\def\sQ{{\widetilde{Q}}}
\def\sU{{\widetilde{U}}}
\def\sD{{\widetilde{D}}}

\def\Hbc{H^\pm bc}
\def\bs{\tilde{b}}
\def\gs{\tilde{g}}
\def\ts{\tilde{t}}
\def\cs{\tilde{c}}
\def\us{\tilde{u}}
\def\cut{{\Lambda}}

\def\ov{\overline}

\def\U1em{{U(1)_{\rm em}}}
\def\to{\rightarrow}

\def\sq2{\sqrt{2}}

\def\tanb{\tan\hspace*{-1mm}\beta}

\def\End{\end{document}}

\begin{document}                                                              
\thispagestyle{empty}

\begin{flushright}
{\large hep-ph/0103178}
\end{flushright}
\vspace*{15mm}

\begin{center}
{\bf {\large 
Soft Supersymmetry Breaking, 
Scalar Top-Charm Mixing and Higgs Signatures \\
}}

\vspace*{15mm}
{\sc\large J. L. Diaz-Cruz}\,$^{\rm a}$,~~
{\sc\large Hong-Jian He}\,$^{\rm b}$,~~ 
{\sc\large C.--P. Yuan}\,$^{\rm c}$

\vspace*{4mm}
$^{\rm a}$~Lawrence Berkeley National Laboratory, Berkeley, 
           California 92710, USA
\\[1.7mm]
$^{\rm b}$~The University of Texas at Austin, Austin, Texas 78712, USA
\\[1.7mm]
$^{\rm c}$~Michigan State University, East Lansing, Michigan 48824, USA
\\
\end{center}

\vspace*{75mm}
\begin{abstract}
\hspace*{-0.35cm}
The squark mass-matrix from the soft supersymmetry (SUSY) breaking
sector contains a rich flavor-mixing structure that allows 
$O(1)$ mixings among top- and charm-squarks while being consistent 
with all the existing theoretical and experimental bounds. 
We formulate a {\it minimal} flavor-changing-neutral current scheme 
in which the squark mixings arise from the non-diagonal scalar trilinear
interactions. This feature can be realized in a class of new models 
with a horizontal $U(1)_H$ symmetry which generates realistic 
quark-mass matrices and provides a solution to the SUSY $\mu$-problem. 
Finally, without using the mass-insertion approximation,
we analyze SUSY radiative corrections to the $H^\pm bc$ and $h^0tc$ couplings,
and show that these couplings can reveal exciting new discovery channels 
for the Higgs boson signals at the Tevatron and the LHC. \\[4mm]
{PACS numbers: \,12.60.-i,\,12.15.-y,\,11.15.Ex}  
\hfill   ~~ 
\end{abstract}

\newpage
\setcounter{page}{1}
\setcounter{footnote}{0}
\renewcommand{\thefootnote}{\arabic{footnote}}

\noindent
{\bf {\large 1. Introduction}} 
\vspace*{3mm}

Weak-scale supersymmetry (SUSY) \cite{review},
as a leading candidate for new physics beyond the standard model (SM),
sensibly explains electroweak symmetry
breaking, but leaves the understanding of
flavor sector as a major challenge.
Being a new fundamental
space-time symmetry between fermions and bosons,
SUSY necessarily extends the SM flavor structure to include  
superpartners for all fermions and thus {\it adds} 
further puzzles to the flavor physics.  
To be consistent with the experimental data,
supersymmetry has to be broken.
The breaking of the SUSY lifts the mass spectrum of the superpartners and
is parametrized by the soft-breaking Lagrangian of the 
Minimal Supersymmetric SM (MSSM) with 
a large number of free parameters. The soft breaking sector 
is often problematic with low-energy flavor changing neutral current 
(FCNC) data without making specific assumptions 
about its free parameters.
One of the most popular assumptions is the proportionality
of the scalar trilinear $A$-terms to the fermion Yukawa couplings.
This is however not a generic feature from the
string-theory constructions, and certain forms of
non-diagonal $A$-terms and their interesting
implications were studied recently \cite{nondA,Mura}.

In this work, 
we focus on the flavor-mixings of three-family squarks which originate 
from the scalar mass and the trilinear interaction terms.
We observe that the current data mainly suppress the FCNCs associated
with the first two family squarks (and in some cases, with the first and third
families), but allow the flavor-mixings between the second- and
third-family squarks, the scharm ($\cs$) and stop ($\ts$),
to be as large as $O(1)$~\cite{FCNC}.
Furthermore, the $O(1)$ $\ts -\cs$ mixings arising from the non-diagonal
$A$-term are consistent with the theoretical bounds derived from 
avoiding the charge-color breaking (CCB) as well as maintaining the
vacuum stability (VS) \cite{CCBVS}. 
Taking a bottom-up approach, we
first formulate a {\it minimal} SUSY FCNC scenario, 
called the Type-A models, in which 
all the observable FCNC effects come from
the non-diagonal trilinear $A$-term
in the $\cs - \ts$ sector.   Then, using the simplest horizontal 
$U(1)_H$ symmetry ({\it \`{a} la} Froggatt-Nielsen\,\cite{FN}),
we construct a class of new models, 
called the Type-B models which not only exhibit  
similar flavor-mixings in the $\ts-\cs$ sector but also
generate realistic quark-mass/mixing pattern and
provide a solution to the SUSY $\mu$-problem.
Such minimal FCNC schemes can reduce 
the general $6\times 6$ squark-mass-matrix down to a $4\times 4$ or
$3\times 3$ matrix involving only the $\cs - \ts$ sector, and
therefore simplify
the exact squark mass-diagonalization and rotations  
without invoking the crude mass-insertion approximation.  
This allows quantitative understanding and predictions of the relevant
FCNC signatures from the squark sector, and thus provides reliable 
probes to the fundamental SUSY flavor structure encoded 
in the soft-breaking Lagrangian.

As we will show, 
exploring the SUSY flavor sector is also important
for revealing exciting new discovery signatures 
from the weak-scale supersymmetry, in addition to probing the
mechanism of soft SUSY breaking.
Applying the minimal FCNC schemes, we analyze the  
SUSY radiative corrections to the $H^\pm bc$ coupling 
and show that this flavor-mixing
coupling can be significant to provide new discovery signals 
of charged Higgs via charm-bottom fusion\,\cite{HY} 
at the on-going Fermilab Tevatron Collider and 
the CERN Large Hadron Collider (LHC). 
Then, we further study the flavor 
changing top-decays into the charm-quark and 
lightest neutral Higgs boson in our schemes, 
and their observability at the LHC.

\vspace*{6mm}
\noindent
{\bf {\large 2. Minimal Supersymmetric FCNC Models 
}}  

\vspace*{3mm}
\noindent
{\bf 2.1. Type-A Minimal SUSY FCNC Models
          with Non-Diagonal $A$-Term}  
\vspace*{3mm}

The MSSM soft-breaking squark-sector contains the following
quadratic mass-terms and trilinear $A$-terms:
\beq
\bea{l}
-\sQ_i^\dag (M_{\sQ}^{2})_{ij}\sQ_j
-\sU_i^\dag (M_{\sU}^{2})_{ij}\sU_j
-\sD_i^\dag (M_{\sD}^{2})_{ij}\sD_j  
+( A_{u}^{ij}\sQ_i H_u\sU_j 
  -A_{d}^{ij}\sQ_i H_d\sD_j + {\rm c.c.} )\,,
\eea
\label{eq:A-term}
\eeq
with $i,j=1,2,3$ being the family indices. 
This gives a generic $6\times6$ mass matrix,
\beq
\sM^2_u =\left\lgroup 
	 \bea{ll}
          M_{LL}^2         &  M_{LR}^2\\[1.5mm]
          M_{LR}^{2\,\dag}   &  M_{RR}^2
	 \eea
         \right\rgroup ,
\label{eq:MU6x6}
\eeq
in the up-squark sector, where
\beq
\bea{ll}
M_{LL}^2 &= M_{\sQ}^2+M_u^2+\f{1}{6}\cos2\beta \,(4m_w^2-m_z^2)\,, \\[2mm]
M_{RR}^2 &= M_{\sU}^2+M_u^2+\f{2}{3}\cos2\beta\sin^2\theta_w\, m_z^2\,,  \\[1mm]
M_{LR}^2 &= \dis A_u v\,\sB/\sqrt{2}-M_u\,\mu\,\cotB \,,
\eea 
\label{eq:MU3x3}
\eeq
with $m_{w,z}$ the masses of $(W^\pm,\,Z^0)$ and $M_u$ the up-quark
mass matrix. For convenience, we will choose the 
super Cabibbo-Kobayashi-Maskawa (CKM) basis for squarks so that
in Eq.\,(\ref{eq:MU3x3}),
$A_u$ is replaced by  $A_u' = K_{UL} A_u K_{UR}^\dag$ and 
$M_u$ by $M_u^{\rm diag}$, etc, with $K_{UL,R}$ the rotation matrices
for diagonalizing $M_u$ to $M_u^{\rm diag}$.
In our {\it minimal} Type-A scheme, we consider all large FCNCs to 
{\it solely} come from non-diagonal $A_u'$ in the up-sector, and those in
the down-sector to be negligible, i.e., we define, at the weak scale,
\beq
A_u' =
      \left\lgroup
      \bea{ccc}
      0 & 0 & 0\\
      0 & 0 & x\\
      0 & y & 1
      \eea
      \right\rgroup A \,,
\label{eq:Au}
\eeq
where, $x$ and $y$ can be of $O(1)$,
representing a naturally large flavor-mixing in the 
$\ts - \cs$ sector. Such a minimal scheme of SUSY FCNC is compelling as 
it is fully consistent with the stringent CCB/VS theory bounds \cite{CCBVS}
as well as the existing data \cite{FCNC}. 
(In the case that the CKM matrix is generated from the down-quark sector 
alone\,\cite{Mura}, $A_u'$ simply reduces back to $A_u$.)
Similar pattern may be also
defined for $A_d$ in the down-sector, but the strong CCB/VS bounds
permit  $O(1)$ $\tilde{b}-\tilde{s}$ mixings only for very large $\tanB$ 
because $m_b \ll m_t$. Thus, to allow a full range of $\tanB$ value, 
we consider an almost diagonal $A_d$.
Moreover, 
indentifying the non-diagonal $A_u$ as the only source of
the observable FCNC phenomena in the Type-A models implies
the squark-mass-matrices $M_{\sQ,\sU}^2$ 
in Eqs.\,(\ref{eq:MU6x6})-(\ref{eq:MU3x3})
to be nearly diagonal.  For simplicity, we define 
\beq
M_{LL}^2 \,\simeq\, M_{RR}^2\, \simeq\,\sm0^2\,{\bf I}_{3\times3}\,,
\label{eq:Degen}
\eeq
with $\sm0$ a common scale of scalar-masses\,\cite{seiberg}.

Within this minimal Type-A scheme, we observe that the first family squarks
$\tilde{u}_{L,R}$ decouple from the rest in (\ref{eq:MU6x6}) 
so that the $6\times6$ mass-matrix is reduced to $4\times4$,
\beq
\sM_{ct}^2  =  
      \left\lgroup
      \bea{cccc}
      ~{\ms}_{0}^2~  &   0             &   0            &   A_x\\[1mm]
      0              &   ~\ms_{0}^2~   &   A_y~         &   0  \\[1mm]
      0              &   A_y~          &  ~\ms_{0}^2~   &   X_t\\[1mm]
      A_x~           &   0             &   X_t~         &  ~\ms_{0}^2
      \eea
      \right\rgroup 
\label{eq:Mct4x4}
\eeq
for squarks $(\cs_L,\,\cs_R,\,\ts_L,\,\ts_R)$, where
\beq
\bea{l}
A_x = x\Ah\,,~~~A_y = y\Ah\,,~~~\Ah = Av\,\sB/\sqrt{2}\,,~~~  
X_t = \Ah - \mu\,m_t\,\cotB \,.                     \\ 
\eea
\label{eq:MctADD}
\eeq
In Eq.\,(\ref{eq:Mct4x4}), we ignore tiny terms of $O(m_c)$ or
smaller. The reduced squark mass matrix (\ref{eq:Mct4x4}) 
has 6 zero-entries in total and 
is simple enough to allow an exact diagonalization.
Especially, for the cases of  
(i)   $x\neq 0,~y=0$, (called Type-A1) and
(ii)  $x=0,~y\neq 0$, (called Type-A2),
one more squark, $\cs_R$ (in Type-A1) or $\cs_L$ (in Type-A2), 
decouples, and the Eq.\,(\ref{eq:Mct4x4}) further
reduces to a $3\times3$ matrix.
We have worked out the general diagonalization of $4\times4$ matrix 
(\ref{eq:Mct4x4}) for any $(x,\,y)$.
The mass-eigenvalues of 
the eigenstates  $(\cs_1,\,\cs_2,\,\ts_1,\,\ts_2)$ are,
\beq
\bea{ll}
M_{\cs1,2}^2 & = \sm0^2 \mp\f{1}{2}|\sqrt{\omega_+}-\sqrt{\omega_-}|\,, \\[2mm]
M_{\ts1,2}^2 & = \sm0^2 \mp\f{1}{2}|\sqrt{\omega_+}+\sqrt{\omega_-}|\,,
\eea
\label{eq:Mass}
\eeq
where $~\omega_\pm = X_t^2+(A_x\pm A_y)^2\,$. From (\ref{eq:Mass}), 
we can deduce the mass-spectrum
of the stop-scharm sector as 
\beq
M_{\ts1} < M_{\cs1} < M_{\cs2} < M_{\ts2} \,.
\label{eq:Mspectrum}
\eeq
In Eq.\,(\ref{eq:Mass}), the stop $\ts_1$ can be
as light as about $100-300$\,GeV for the typical range of 
$\sm0\gtrsim 0.5-1$\,TeV.
The $4\times4$ rotation matrix of the diagonalization
is given by, 
\beq
\bea{l}
\left\lgroup
\bea{l}
\cs_L\\
\cs_R\\
\ts_L\\
\ts_R
\eea
\right\rgroup
\!\!=\!\!
\left\lgroup
\bea{rrrr}
 c_1c_3  &  c_1s_3  & s_1s_4  &  s_1c_4  \\
-c_2s_3  &  c_2c_3  & s_2c_4  & -s_2s_4  \\
-s_1c_3  & -s_1s_3  & c_1s_4  &  c_1c_4  \\
 s_2s_3  & -s_2c_3  & c_2c_4  & -c_2s_4
\eea
\right\rgroup
\!\!
\left\lgroup
\bea{l}
\cs_1\\
\cs_2\\
\ts_1\\
\ts_2
\eea
\right\rgroup , 
\eea
\label{eq:rotation}
\eeq
with
\beq ~~~
s_{1,2}=\dis \f{1}{\qs} 
\[1-\f{X_t^2\mp A_x^2\pm A_y^2}{\sqrt{~\omega_+\omega_-}} \]^{1/2}\!,~~~
s_4
=\f{1}{\qs}\,,
\label{eq:rotation2}
\eeq
and $s_3=0$ (if $xy=0$), or, $s_3=1/\qs$
(if $xy\neq 0$), where $s_j^2+c_j^2=1$.

\vspace*{5mm}
\noindent
{\bf 2.2. Type-B Minimal SUSY FCNC Models
          with a Horizontal $U(1)$ Symmetry} 
\vspace*{3mm}

The minimal Type-A SUSY FCNC schemes with a non-diagonal $A_u$-term,
as discussed above,  are truly economical as 
they uniquely result from imposing all the stringent 
theoretical and experimental bounds. 
Below, we further support 
such type of FCNC in the $\ts-\cs$ sector by providing 
theoretically compelling constructions 
based upon a minimal family symmetry.
An attractive approach is to make use of the simplest
horizontal $U(1)_H$ symmetry to generate realistic flavor
structure of both quarks and squarks,
via proper powers of a single suppression factor\,\cite{FN,U1H},
and to provide a solution to the SUSY $\mu$-problem.
For convenience, we define the suppression factor
$\ep=\langle S\rangle/\cut$ 
to have a similar size as the Wolfenstein-parameter $\la$ in
the CKM matrix, 
i.e., $\ep\simeq \la \simeq 0.22$\,\cite{U1H}.
Here, $\langle S\rangle$ is the vacuum expectation
value of a singlet scalar $S$, responsible for 
spontaneous $U(1)_H$ breaking,
and $\Lambda$ is the scale at which the $U(1)_H$ breaking is 
mediated to light fermions.
In general, the supermultiplets of three-family quarks/squarks 
may carry different $U(1)_H$ charges, as defined in Table\,1.
\\ 

\noindent
{Table\,1. 
Quantum number assignments under the horizontal symmetry $U(1)_H$.
}
\begin{center}
\begin{tabular}{ccc|ccc|ccc|ccc}
\hline\hline
&&&&&&&&&&&\\[-2.5mm]
$\,Q_1$      & $~Q_2$      & $~Q_3$      \, &   
$~\ov{u}_1$ & $~\ov{u}_2$ & $~\ov{u}_3$ \, &
$~\ov{d}_1$ & $~\ov{d}_2$ & $~\ov{d}_3$ \, & 
$~H_u$      & $~H_d$      & $~S$        \, \\ [1.5mm] 
\hline
&&&&&&&&&&&\\[-2.5mm]
$\,h_1$     & $h_2$      & $h_3$      \, &
$\alpha_1$ & $\alpha_2$ & $\alpha_3$ \, &
$\beta_1$  & $\beta_2$  & $\beta_3$  \, &
$\xi$      & $\xi'$    & $-1$       \\[1.5mm]
\hline\hline
\end{tabular}
\end{center}
\vspace*{4mm}

It is then straightforward to deduce the following hierarchy structures
in the quark mass matrices:
\beq
M_u^{ij}\sim\dis\f{v_u}{\qs}\la^{\alpha_i+h_j+\xi}\,,~~
M_d^{ij}\sim\dis\f{v_u}{\qs}\f{1}{\tanB}\la^{\beta_i+h_j+\xi'}\,.
\label{eq:MuMd}
\eeq
Similarly, in the CKM matrix,
\beq
\(V_{us},\,V_{cb},\,V_{ub}\) \sim 
\(\la^{h_1-h_2},\,\la^{h_2-h_3},\,\la^{h_1-h_3}\)\,.
\label{eq:CKM}
\eeq
Unlike Ref.\,\cite{U1H}, the key ingredient of our model-buildings
is to impose a new condition,
\beq
\alpha_2\,=\,\alpha_3 \,,
\label{eq:key}
\eeq
which ensures the mixings between $\cs$ and $\ts$ in the
squark mass matrix to be naturally of  $O(1)$.  From the 
condition (\ref{eq:key}) and
the current data of quark-masses and CKM angles (which can all be
counted in powers of $\la$),  we find an almost unique
solution for all quark/squark quantum numbers (cf.,\,Table\,2),
which is  based upon a single $U(1)_H$ symmetry and will be called
the {\it minimal} Type-B scheme hereafter.   In Table\,2, we
consider $\tanB \sim O(1)$ for the down-type quark mass-matrix
$M_d$ [cf. Eq.\,(\ref{eq:MuMd})].      The extension to 
a larger $\tanB$ only affects the quantum numbers of $\ov{d}_j$'s 
in a trivial way as it just contributes an overall factor 
$1/\tanB\sim\la^k$ (with the integer $k\sim 0.66\log\tanB$) to $M_d$ 
in Eq.\,(\ref{eq:MuMd})
and thus simply adds $-k$ to each quantum number of
$\ov{d}_j$ listed in Table\,2.
\\[2mm]

\noindent
{Table\,2.\,Quantum number assignments are derived for the minimal Type-B model
with $\tanB \sim O(1)$.  
}
\vspace*{-1mm}
\begin{center}
\begin{tabular}{ccc|ccc|ccc|ccc}
\hline\hline
&&&&&&&&&&&\\[-2.5mm]
$Q_1$      & $Q_2$      & $Q_3$            &   
$~\ov{u}_1$ & $~\ov{u}_2$ & $~\ov{u}_3$    &
$~\ov{d}_1$ & $~\ov{d}_2$ & $~\ov{d}_3$    & 
$~H_u$      & $~H_d$      & $~S$           \\ [1.5mm] 
\hline
&&&&&&&&&&&\\[-2.5mm]
$4$          & $3$           & $0$            &
$3\!-\!\xi$  & $-\xi$        & $-\xi$         &
$4\!-\!\xi'$ & $3\!-\!\xi'$  & $3\!-\!\xi'$   &
$\xi$        & $\xi'$   & $-1$           \\[1.5mm]
\hline\hline
\end{tabular}
\end{center}
\vspace*{3mm}

There are some slight variations of this minimal Type-B model, by allowing 
the quantum numbers of $Q_j$'s to have $\xi(\xi')$-dependence,
but they all predict the same patterns for the quark masses and mixings.
By this construction, we have attempted to simultaneously solve the
SUSY $\mu$-problem from the same $U(1)_H$.
A dynamical $\mu$-term can originate from 
$\,(\kappa/\cut^{n-1}) S^nH_uH_d\,$, with $n=\xi+\xi'$, such that 
$\,\mu=\kappa \la^{n-1}\langle S\rangle\,$ 
is generated at a scale 
$\langle S\rangle \ll M_{\rm Planck}$.
Hence, a weak-scale value of $\mu$ can be obtained 
by properly choosing $n$ for a given  
$\langle S\rangle$.
If $U(1)_H$ is not related to the SUSY $\mu$-problem, 
the minimal Type-B model becomes truly unique,
corresponding to a special case of $\xi=\xi'=0$ in Table\,2.
With Table\,2, we can readily derive
the structures of quark/squark mass-matrices. 
For instance, the up-quark mass-matrix takes the form of 
\beq
M_u \,\sim\, \dis \f{v_u}{\qs}
\left\lgroup
            \bea{ccc}
            \la^7  &  \la^4  &  \la^4 \\
            \la^6  &  \la^3  &  \la^3 \\
            \la^3  &   1     &   1    
            \eea
\right\rgroup ,
\label{eq:MuAupower}
\eeq
for any $\tanB\gtrsim 1$, 
while the squark mass-matrices $M_{LL}^2$ and $M_{RR}^2$ in
Eq.\,(\ref{eq:MU6x6}) are deduced as,
\beq
\ba{l}
M_{LL}^2
\,\sim\, \sm0^2\!
\left\lgroup
       \bea{ccc}
      1    &  \la   &  \la^4  \\
     \la   &   1    &  \la^3  \\
     \la^4 &  \la^3 &    1   \\ 
       \eea
\right\rgroup ,~~~~~~~
M_{RR}^2
\,\sim\, \sm0^2\!
\left\lgroup
       \bea{ccc}

         1   &  \la^3    &  \la^3  \\
      \la^3  &    1      &    1      \\
      \la^3  &    1      &    1    \\
       \eea
\right\rgroup \!.
\ea
\label{eq:MLLMRR}
\eeq
Eqs.\,(\ref{eq:MuAupower}) and (\ref{eq:MLLMRR}) 
show a partial quark-squark ``alignment'' that 
effectively suppresses the FCNCs between 1st and 2nd(3rd) families, 
while at the same time provides  $O(1)$
mixings in the $\ts-\cs$ sector of $M_{RR}^2$.
Furthermore, because squarks carries the same $U(1)_H$-charge as quarks,
the $\ts-\cs$ mixings originating from a non-diagonal $A_u$ term
is {\it predicted} to share the same hierarchy structure as that
in  $M_u$ [cf. Eq.\,(\ref{eq:MuAupower})].
This, however, does not imply an exact ``proportionality'' between
$A_u$ and $M_u$ because the power-counting of $\la$  
allows the coefficients in the corresponding entries of 
$A_u$ and $M_u$ to differ by ratios of $O(1)$.
Ignoring the small $O(\la^3)\simeq 1\%$ terms in $M_u$, we can
diagonalize $M_u$ by a $2\times2$ rotation of the singlet
quarks $(\ov{c},\,\ov{t})$.
Under the super-CKM basis, the off-diagonal block $M_{LR}^2$ of 
the Eq.\,(\ref{eq:MU6x6}) becomes 
\vspace*{-1mm}
\beq
M_{LR}^2 \,=\, \dis A_u' v\,\sB/\sqrt{2}-M_u^{\rm diag}\,\mu\,\cotB \,,
\eeq
where
$A_u' = K_{UL}A_uK_{UR}^\dag=A_uK_{UR}^\dag +O(\la^3)$ and the 
singlet-quark rotation matrix $K_{UR}$ contains a nontrivial
sub-matrix involving only the second and the third family squarks. 
After neglecting the tiny $O(\la^3)$ terms,
we can parametrize the {\it minimal} $A_u'$ term of the Type-B 
model as 
\vspace*{-1mm}
\beq
\bea{l}
A_u' =
      \left\lgroup
      \bea{ccc}
      0 & 0 & 0\\
      0 & 0 & 0\\
      0 & y & 1
      \eea
      \right\rgroup   A \,, \\[-1.5mm]
\eea
\label{eq:Aup}
\eeq
where the size of the mixing parameter $y$ is naturally of $O(1)$.
Thus, under this construction, the squarks $(\us_{L},\,\us_R,\,\cs_L)$
decouple and the Eq.\,(\ref{eq:MU6x6})
greatly reduces to a $3\times3$ matrix, which takes the form,
under the basis $(\cs_R,\,\ts_L,\,\ts_R)$,
\vspace*{-1mm}
\beq
\widetilde{M}_{ct}^2[{\rm B}] =
\left\lgroup
            \bea{ccc}
            \ms_{0}^2\,  &   A_y    \,    &    x\ms_{0}^2  \\[1.5mm]
             A_y    \,   &  \ms_{0}^2\,   &    X_t         \\[1.5mm]
           x\ms_{0}^2\,  &   X_t      \,  &   \ms_{0}^2   
            \eea
\right\rgroup .
\label{eq:MctB}
\eeq 
In the above equation, $A_y=yAv\,\sB/\sqrt{2}$, and the parameter 
$x=O(1)$ characterizes the mixing of $\cs_R-\ts_R$
in the mass matrix $M_{RR}^2$ [cf., Eq.\,(\ref{eq:MLLMRR})].
For convenience, we define the typical case 
with $y=0$ as the Type-B1 scheme, and 
that with $x=0$ as the Type-B2 scheme.
We note that the Type-B2 model is {\it identical} to the 
Type-A2 model as they have the same form of the non-diagonal 
$A_u$ and $\cs_L$ decouples. 
Also, Type-B2 mass matrix with $y=0$ in Eq.\,(\ref{eq:MctB}) takes 
the same structure as that of Type-A1 
except that Eq.\,(\ref{eq:MctB}) involves $\cs_R$ (not $\cs_L$) 
and its $x$ originates from $M_{RR}^2$ (not $A_u$). 
Without losing generality, we will study the physical applications of
the typical Type-A1 and -A2 (-B2) models in the following sections.

\vspace*{7mm}
\noindent
{\bf {\large 
3. SUSY Radiative Corrections and Higgs Signatures at Colliders
}} 

\vspace*{3mm}
\noindent
{\bf 3.1. SUSY Induced $H^\pm bc$ Vertex
          and $H^\pm$ Production at Hadron Colliders} 
\vspace*{2mm}

Different from the TopColor models and 
the Type-III of two-Higgs-doublet models 
discussed in Ref.\,\cite{HY}, the MSSM Higgs sector has no
FCNC at the tree level and the same is true for flavor-changing mixings
in the charged sector (except the usual CKM mechanism).
Thus, the new flavor-changing effects in the neutral and charged sectors
of the MSSM have to be generated radiatively.   From 
Eq.\,(\ref{eq:rotation}) and its resulted Feynman rules, 
we can compute the dominant SUSY-QCD 
radiative corrections to the $\Hbc$ coupling.
It contains vertex corrections 
[scharm(stop)-sbottom-gluino loop]
and self-energy corrections 
[scharm(stop)-gluino loop], and can be summarized as,
\beq
\bea{rl}
\dis 
{\Gamma}_{H^+b\ov{c}} & =~ 
i\,\ov{u}_c(k_2)\(F_LP_L+F_RP_R\)u_b(k_1) \,,\\[2mm]
\dis 
F_{L,R} & =~ F^{0}_{L,R} + F^{V}_{L,R} + F^{S}_{L,R} \,, 
\eea
\label{eq:Hbc}
\eeq 
where $P_{L,R}=(1\mp\gamma_5)/2$ and
the tree-level results are
\beq
(F_L^0,\,F_R^0)
= \dis \f{gV_{cb}}{\qs m_w}
\( m_c\cotB,\, m_b\tanB \)  \,,
\label{eq:FLRTree}
\eeq
with $V_{cb}$ being the CKM mixing matrix element involving
$c$ and $b$ quarks.
The one-loop vertex corrections in the Type-A1 model give, 
\beqn
F_L^V & \!=\! & 0\,,
\nonumber 
\\[-2mm]
\label{eq:FVER}  
\\[-2mm]
\dis F^{V}_{R} 
        & \!=\! & \frac{\alpha_s}{3 \pi}
                  m_{\tilde{g}}
         \!\sum_{j,k}\!
            \kappa^R_{jk}
         C_0(m_H^2,0,0; m_{\bs_j}, m_{\gs}, m_{\tilde{u}_{k}})\,, 
\nonumber 
\eeqn
where 
$\tilde{u}_{k}\in (\tilde{c}_{2},\tilde{t}_1,\tilde{t}_2)$,
$\tilde{b}_j\in (\tilde{b}_1,\tilde{b}_2)$, and
$C_0$ is the 3-point $C$-function of Passarino-Veltman\,\cite{PV}.
$\kappa^R_{jk}$ is the product of the relevant
$H^\pm$-$\bs_j$-$\us_{k}$, $\bs_j$-$\gs$-$b$ and $\us_{k}$-$\gs$-$c$ couplings.
The $\bs_L-\bs_R$ mixings are also included, which can be sizable 
for large $\tanB$.  Furthermore, the Type-A1 self-energy corrections
yield   
\beqn
F^S_L &=& 0 \,,
\nonumber 
\\[-2mm]
\label{eq:FSEL}
\\[-2mm]
F^{S}_{R} 
        &=& \widehat{F}^0_R \frac{\alpha_s s_1}{3\pi}
                 \f{m_{\tilde{g}}}{m_t} 
           \!\sum_{j=1,2}(-)^{j+1} \!
           B_0 ( 0; m_{\gs}, m_{\ts_j} ),  
\nonumber 
\eeqn
where $B_0$ is the 2-point Passarino-Veltman function \cite{PV} and 
$s_1$ is given in Eq.\,(\ref{eq:rotation}) for $y=0$, and  
tree-level $H^\pm tb$ couplings
$(\widehat{F}^0_L,\,\widehat{F}^0_R)
=(gV_{tb}/\sqrt{2}m_w)(m_t\cotB,\,m_b\tanB)$.
In  Eqs.\,(\ref{eq:FVER})-(\ref{eq:FSEL}),
the tiny sub-leading terms suppressed by the powers of 
$m_{c}/m_{t,\gs}$ have been ignored.

The form factors $F^{V,S}_{L}$ and $F^{V,S}_{R}$
for the Type-A2 and Type-B2
models can be easily obtained
from Eqs.\,(\ref{eq:FVER})-(\ref{eq:FSEL})
by the exchanges of  $L \leftrightarrow R$ and $x \to y$.
We note that $F_{L}\,(F_R)$  vanishes in
Type-A1 (Type-A2 $\&$ -B2) schemes because $\cs_R$ ($\cs_L$) decouples.

We find that the effective $\Hbc$ couplings $F_{L,R}$ are 
typically around $0.03-0.1$ for $(x,\,y)\simeq 0.5-0.9$, 
$(A,\,\ms_0 )\simeq 0.5-2$\,TeV, and $\tanB \simeq 15-50$.
Therefore, the production of a 
charged Higgs boson via the $b$-$c$
quark fusion process can become important in a wide range of
the parameter space.
In Fig.\,1(a), the production cross sections
of $H^\pm$ via $p\bar{p}/pp\to H^\pm X$ 
at the Tevatron and LHC are shown for  
$(\mu,\,m_{\tilde g},\,\sm0)=(300,\,300,\,600)$\,GeV,
$(A,-A_b)=1.5$\,TeV,
$\tanB=(15,\,50)$, and $x=0.75$ 
in the Type-A1 model.
The dotted and dash curves are the SM production rates 
of $cs\to H^\pm$ and $cb\to H^\pm$ (induced by 
$V_{cb}\simeq 0.04$), respectively, 
after including the complete next-to-leading order (NLO) SM-QCD
corrections\,\cite{NLO}. (The NLO SM-QCD corrections include the
subprocesses with one single gluon in the initial state.)
The solid curve is the full  $cb\to H^\pm$  production rate
as a function of the Higgs boson mass after including both
the SM-QCD and new SUSY-QCD corrections.
As indicated, the SUSY loop corrections can significantly 
dominate over the CKM-suppressed $F^0_{L,R}$ 
contributions by a factor of $\sim\!2-5$. 
In Fig.\,1(b) and (c), we show the $K$-factor, defined as
the ratio of $(F_L^2+F_R^2)$ over $({F_L^0}^2+{F_R^0}^2)$, which 
characterizes the enhancement of the $H^\pm$ production 
rate by the SUSY loop contributions.

To study the detection of a $H^\pm$ scalar, we 
consider its decay modes in the following.
For $m_H\lesssim\!190$\,GeV, the $\tau\nu$ channel dominates, 
and for $m_H\gtrsim 190$\,GeV, $H^\pm$ mostly decays into 
the $tb$ channel unless $H^\pm$ mass is above the threshold of
$W^\pm h^0$, in which case the $W^\pm h^0$ channel 
(with $W\to \ell\nu$ and $h^0\to b\bar{b}$) 
can become important as well\,\cite{NLO}. 
Fig.\,1 suggests that the Tevatron may be sensitive to 
the $H^\pm$ signals for $m_H$ below $\sim\!300$\,GeV, with 
an integrated luminosity of $2-20$\,fb$^{-1}$  per detector, while
the LHC can potentially probe the full mass-range of $H^\pm$ 
with an integrated luminosity of about $100$\,fb$^{-1}$.
A detailed Monte Carlo simulation is needed to
further quantify the discovery potential of the Tevatron and the LHC,
which is beyond the scope of this Letter.

We have also examined the similar SUSY-induced enhancement $K$-factor 
for the $H^\pm$ production rate in the type-A2 and -B2 models
and found that in the low $\tan \beta$ ($\lesssim 5-10$) region it 
can reach to about $2-5$ for $y=0.5-0.9$.
But, as $\tan\beta$ increases to above $\sim\! 15$, the 
enhancement factor decreases to less than 
$\sim 1.3$ for Higgs mass below 1\,TeV.

\vspace*{5mm}
\noindent
{\bf 3.2. SUSY Induced $h^0tc$ Vertex and
          Neutral Higgs Signal from Top Decay }
\vspace*{3mm}

It is known that the SM branching ratio of the flavor-changing top 
decay $t \to c h^0$ is extremely small 
($\lesssim 10^{-13}-10^{-14}$ \cite{SM-tch}), so that
this channel becomes an excellent window 
for probing new physics\,\cite{sola,tch-bound,han}.
Our minimal SUSY FCNC models predict
the branching ratio of the $t \to c h^0$ decay to
be substantially above the SM value so that this decay mode becomes
observable at the LHC.
(Note that this decay channel
is always kinematically allowed in the MSSM.)

In our minimal FCNC schemes,  
the one-loop SUSY QCD induced $tch^0$ coupling can be written as
\beq
\ba{l}
\dis 
{\Gamma}_{t\bar{c}h} \, =\, 
i\,\ov{u}_c(k_2)\(F_LP_L+F_RP_R\)u_t(k_1) \,,\\[4mm]
\dis 
F_{L,R}  \,=\, F^{V}_{L,R} + F^{S}_{L,R} \,, 
\ea
\label{eq:Htc}
\eeq 
which contains the vertex corrections (from
scharm-stop-gluino, stop-stop-gluino and scharm-scharm-gluino, 
triangle loops), and the self-energy corrections (from 
stop-gluino and scharm-gluino loops).
The one-loop vertex corrections in Type-A1 are,
\beqn
\dis 
F^{V}_{L} & \!\!=\!\! & \frac{\alpha_s}{3 \pi}                  
         \!\sum_{j,k}\!
            \lambda^L_{jk} m_t (C_0+C_{11})
           (m_h^2,m_t^2,0; 
            m_{\tilde{u}_{j}}, m_{\gs}, m_{\tilde{u}_{k}})\,, 
\nonumber
\\[-1mm]
\label{eq:FVERch}
\\[-2mm]
\dis
F^{V}_{R} & \!\!=\!\! & \frac{\alpha_s}{3 \pi}                  
         \!\sum_{j,k}\!
            \lambda^R_{jk} m_{\gs} C_0
           (m_h^2,m_t^2,0; 
            m_{\tilde{u}_{j}}, m_{\gs}, m_{\tilde{u}_{k}})\,,
\nonumber
\eeqn
where 
$\tilde{u}_{j,k}\in (\tilde{c}_{2},\tilde{t}_1,\tilde{t}_2)$,
and
$(C_0, C_{11})$ are the 3-point $C$-function of Passarino-Veltman.
$\lambda^{L,R}_{jk}$ is the product of the relevant
$h-\us_j-\us_{k}$ and $\us_{k}-\gs-t(c)$ couplings,
derived from applying the squark-rotation (\ref{eq:rotation}). 
The Type-A1 self-energy corrections yield,
\beqn
F^S_L &\!\!=\!& 0\,,
\nonumber
\\[-2mm]
\label{eq:FSELch}
\\[-1mm]
\dis
F^{S}_{R} 
         &\!\!=\!& \widetilde{F}_0 \frac{\alpha_s s_\theta}{3\pi}
                 \f{m_{\tilde{g}}}{m_t} 
         \[ B_0 ( 0; m_{\gs}, m_{\ts_2}) -
            B_0 ( 0; m_{\gs}, m_{\ts_1})   \] \,,~~~
\nonumber
\eeqn
where $B_0$ is the 2-point Passarino-Veltman function and 
$s_\theta$ is given in Eq.~(\ref{eq:rotation}) with $y=0$. 
$\tilde{F}_0$ denotes the tree-level $h^0-t-\bar{t}$ coupling,
and is given by
$\widetilde{F}^0 = (m_t/v) (\cos\alpha /\sin\beta ) $.
Again, in  Eqs.\,(\ref{eq:FVERch})-(\ref{eq:FSELch}), 
we have ignored
the tiny sub-leading terms suppressed by the powers of 
$m_{c}/m_{t,\gs}$. 

The form factors $F^{V,S}_{L,R}$ in the Type-A2 model 
can be obtained
from  (\ref{eq:FVERch})-(\ref{eq:FSELch})
by the exchanges of $L\leftrightarrow R$ and $x\to y$ everywhere.

For the numerical study, we assume that the only dominant
decay mode of the top quark is its SM decay mode, $t \to bW$.
Thus, the decay branching ratio of $t \to c h^0$ is given by, 
Br$[t\to ch^0]\simeq \Gamma [t \to ch^0]/\Gamma [t \to bW]$,
with the partial decay width  
\beq 
\Gamma (t \to ch) ~=~ \f{m_t}{16 \pi} 
                    \[ 1- \f{m^2_h}{m^2_t} \]^{\f{1}{2}}    
                   \( F_L^2 + F_R^2  \)\,, 
\label{eq:Widtch}
\eeq 
where the form factors $(F_L,\,F_R)$ are defined in Eq.\,(\ref{eq:Htc}).
As summarized by Table\,3, 
in our minimal SUSY-FCNC schemes, the decay branching ratio 
${\rm Br}[t\to ch^0]$
can be as large as $10^{-3} - 10^{-5}$ over
a large part of the SUSY parameter space where
the mass of the lightest Higgs boson $h^0$ 
is around $110-130$\,GeV.
We see that these  decay branching ratios are very sensitive
to the mixing parameter $x$ when it varies from 0.5 to 0.9.
One reason is that the branching ratio (or decay width) contains,
besides other mass-diagonalization effects, a sensitive overall 
power factor $x^2$ which comes from the
stop-scharm mixing induced flavor-changing coupling in the
squark-squark-gluino triangle loop 
and squark-gluino self-energy loop.
Another reason is that unlike the usual analyses with mass-insertion
approximation, we have performed exact squark mass
diagonalization [cf. Eqs.\,(8)-(11)],
so that stops and/or scharms can have significant
mass-splittings. 
For instance,  Eq.\,(8) shows that the mass-splitting 
between two top-squarks is not just due to the usual
left-right mixing from $X_t$, but also arises from the non-diagonal
$A$-term, $xA$ (and $yA$). The latter further enhances the
mass-splittings and results in a light $\tilde{t}_1$ of mass
$\sim\!300-100$\,GeV for $x=0.5-0.9$, and the heavier $\tilde{t}_2$
always has mass above the input $\widetilde{m}_0$ 
(set as 600\,GeV in Table\,3). 
The radiative vertex corrections to $tch^0$ coupling are thus
dominated by the diagram with  
\,$\tilde{t}_1-\tilde{t}_1-\tilde{g}$\, triangle loop
as we have explicitly verified from the  3-point  
Passarino-Veltman $C$-functions in Eq.\,(\ref{eq:FVERch}). 
This second reason further increases the sensitivity of
our decay branching ratios to the mixing parameter $x$.
From Table\,3, we also see that for moderate mixings
with $x\lesssim 0.5$, the branching ratios are generally 
bounded to around the order of $10^{-5}$, consistent with other
studies in the literature \cite{sola}.
The similar conclusion also holds for our
Type-A2 and -B2 models.

Since the LHC with an 
integrated luminosity of $100$\,fb$^{-1}$ can produce
about $10^8$ $t$ and $\bar{t}$ pairs\,\cite{tt-rev},
it can have a great sensitivity to discover this decay channel
and test the model predictions, by demanding one top decaying into the 
usual $bW^\pm$ mode and another to the FCNC $ch^0$ mode. 
As shown in a recent model-independent Monte Carlo 
analysis\,\cite{tch-bound}, the LHC ($100$\,fb$^{-1}$) 
can already measure the Br$[t\to ch^0]$ 
down to the level of $4.5\times 10^{-5}$ at the $95\%$\,C.L. 
The future Linear Collider, with a high luminosity, 
is also expected to have a good sensitivity to detect this channel.

\vspace*{5mm}
\noindent
{Table\,3. 
Br$[\,t\!\to\! c\,h^0\,]\times 10^{3}$
is shown for a sample set of Type-A1 inputs with 
$(\sm0,\mu,A)=(0.6,0.3,1.5)$\,TeV and Higgs mass
$M_{A^0}=0.6$\,TeV.
The three numbers in each entry correspond to
$x = (0.5,\,0.75,\,0.9)$, respectively.
}
\vspace*{1.5mm}
\begin{center}
\begin{tabular}{c||c|c|c}
\hline\hline
&&&\\[-2.5mm]
$m_{\tilde g}$ & $\tanb=5$ & 20 & 50~         \\ [1.5mm] 
\hline
&&&\\[-2.5mm]
~$100$\,GeV~   & (0.011,\,0.10,\,0.81)  &  (0.015,\,0.19,\,4.6) 
             & (0.016,\,0.21,\,7.0)\,   \\[1.5mm]
\hline
&&&\\[-2.5mm]
~$500$\,GeV~ &  (0.011,\,0.09,\,0.41)  &   (0.015,\,0.13,\,1.0) 
           &  (0.016,\,0.14,\,1.2)\,   \\[1.5mm]
\hline\hline
\end{tabular}
\end{center}

\noindent
{\bf {\large 4. Conclusions  }}\\
\vspace*{1.5mm}

The three-family squark mass-matrix originating from 
the soft SUSY breaking sector contains 
a rich flavor-mixing structure. 
In this work, we have constructed the minimal FCNC 
schemes for the squark mass-terms and the scalar 
trilinear interactions which are 
consistent with the existing experimental and theoretical bounds. 
We find that the $O(1)$  large mixings among 
the top- and charm-squarks are allowed.  
We demonstrate that this feature 
can be naturally realized in a class of new models with 
a horizontal $U(1)$ symmetry which 
also generate realistic quark-mass 
pattern and solve  the SUSY $\mu$-problem.     
Finally, we systematically analyze the dominant supersymmetric 
radiative corrections to the $bcH^\pm$ and $tch^0$ couplings
in our minimal schemes, 
without using the mass-insertion approximation. 
We show that these couplings can be significant to provide 
new discovery signatures of the charged and neutral 
Higgs bosons at the Fermilab Tevatron and the CERN LHC.

\vspace*{4mm}
\noindent
{\bf {\large Acknowledgments}}\\[1.7mm]
We thank G. L. Kane, D. A. Dicus, T. Han and Y. Nir for 
valuable discussions and reading the manuscript.
This work was supported by U.S. DOE and NSF.



\begin{figure*}
\vspace*{2mm}  
\begin{center}
\epsfig{file=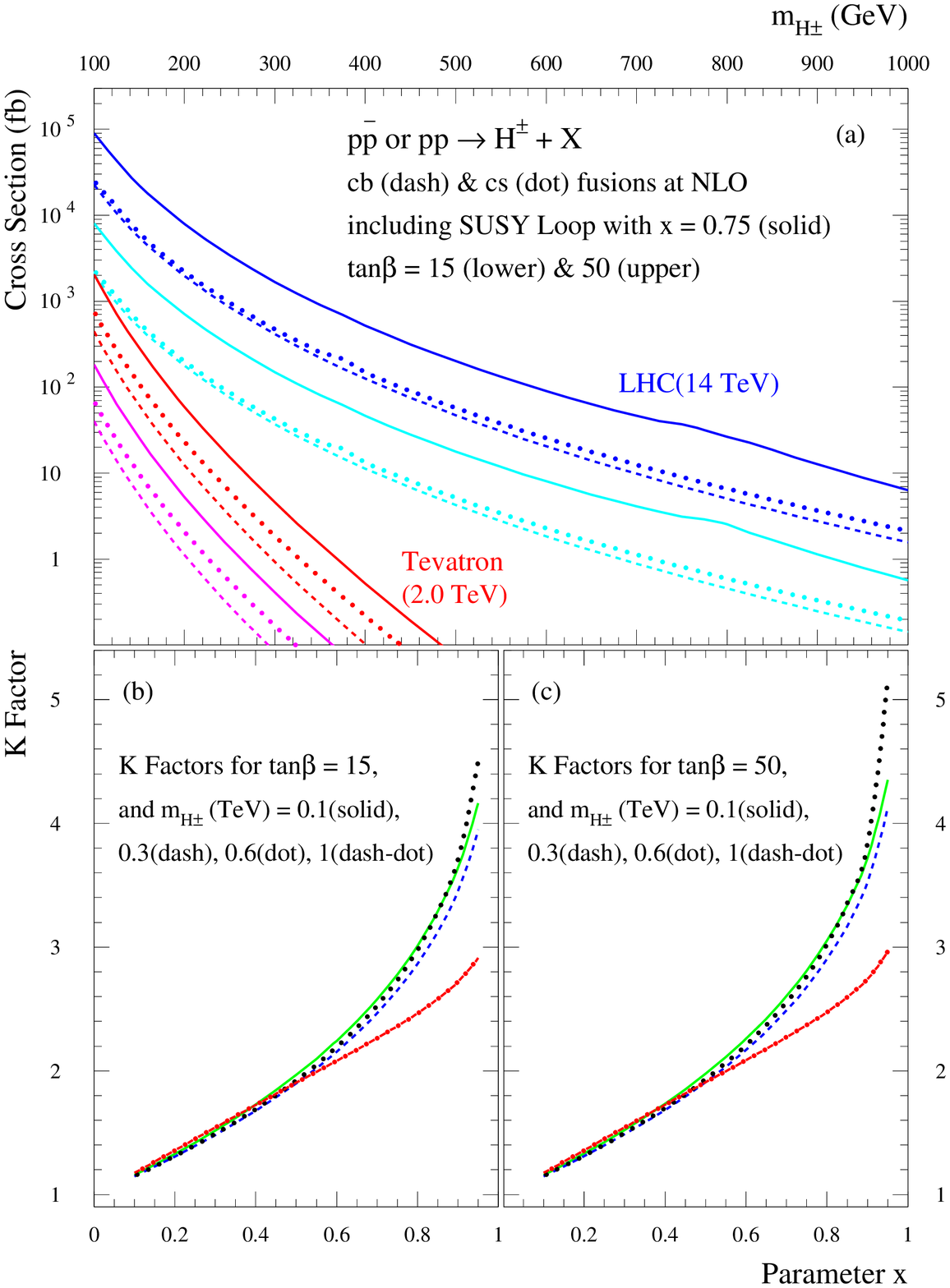,
width=18cm, height=19cm}
\end{center}
\vspace*{-1mm}
\caption{(a). $H^\pm$ production via $cb$ (and $cs$) 
              fusions at hadron colliders,
              with sample inputs $\tanb = 15\,(50)$ 
              shown as lower (upper) set of curves,
              and $x=0.75$.
         (b) and (c). Factor 
              $K\equiv \(F_L^2+F_R^2\)/\({F_L^0}^2+{F_R^0}^2\)$
	      for $H^\pm bc$ vertex,
              as a function of parameter $x$ and for $\tanb=(15,\,50)$.
}
\label{fig:fig1}    
\end{figure*}

\end{document}